\documentclass[prd,aps,twocolumn,floatfix,nofootinbib]{revtex4-1}
\usepackage{amssymb,graphicx}
\usepackage{epsfig}
\usepackage[usenames]{color}

\newcommand{\beqn}{\begin{eqnarray}}
\newcommand{\eeqn}{\end{eqnarray}}
\newcommand{\beq}{\begin{equation}}
\newcommand{\eeq}{\end{equation}}

\begin{document}

\title{Simulating extreme-mass-ratio systems in full general relativity}
\author{William E.\ East and Frans Pretorius}
\affiliation{
Department of Physics, Princeton University, Princeton, New Jersey 08544, USA\\
}

\begin{abstract}
We introduce a new method for numerically evolving the full Einstein field equations 
in situations where the spacetime is dominated by a known background solution.
The technique leverages the knowledge of the background solution to subtract
off its contribution to the truncation error, thereby more efficiently achieving a
desired level of accuracy.
We demonstrate the method by applying it to the radial infall of a solar-type 
star into supermassive black holes with mass ratios $\geq 10^6$. The self-gravity
of the star is thus consistently modeled within the context of general relativity, and
the star's interaction with the black hole computed with moderate computational cost, despite 
the over five orders of magnitude difference in gravitational potential (as defined by the ratio of mass to radius).
We compute the tidal deformation of the star during infall, and the gravitational
wave emission, finding the latter is close to the prediction of the point-particle limit.
\end{abstract}

\maketitle

\section{Introduction}
\label{intro}

In recent years, rapid progress has been made in extending the purview of the field of numerical general relativity 
to a wider class of binary systems.  Numerical solutions of the full Einstein equations
have been used to study not only compact objects of comparable masses, 
but also black hole (BH) binaries with mass ratios of up to 100:1~\cite{Lousto:2010ut,Sperhake:2011ik},
white dwarf-intermediate mass BH systems~\cite{2012ApJ...749..117H}, and
neutron star-pseudo white dwarf
mergers~\cite{Paschalidis:2009zz,Paschalidis:2010dh,Paschalidis:2011ez}.
In the latter cases, the compaction (ratio of mass to radius in geometric 
units, $G=c=1$, which we use throughout) of the white dwarf was $\sim10^{-4}$, and $\sim10^{-2}$ for the pseudo white dwarf.  
Here we are interested in pushing this domain
of study even further to BH-stellar systems where the star has compaction $\sim10^{-6}$, and
the mass ratio reaches upwards of $10^6$:1.
However, simulating these systems with standard methods is very computationally expensive 
due to the disparate scales in the problem. 
In order to accurately recover the dynamics of the system, the truncation error
from evolving the BH must be reduced below the level of the star's 
contribution to the solution.  Since the star's contribution to the spacetime
metric is many orders of magnitude smaller than that of the BH, this
will require exceedingly high resolution compared to the scale that would otherwise 
be set by the BH alone. 
In this paper we introduce a new method for numerically evolving these systems in full
general relativity that makes use of the knowledge of the analytic solution of the larger object in
order to subtract off the truncation error of the background solution.  
This method allows extreme-mass-ratio systems to be simulated more efficiently and with greater
accuracy at a given resolution.

One of the motivations for the development of this method is the study of tidal disruption 
of stars by supermassive BHs.  Considerable interest in these events has been sparked
by the observation in the optical through ultraviolet wavelengths of candidate disruptions and 
subsequent relativistic outflows associated with the fallback of disrupted material onto the supermassive BH
~\cite{1996A&A...309L..35B,1999A&A...349L..45K,2003ApJ...592...42G,2006ApJ...653L..25G,2008ApJ...676..944G,2009A&A...495L...9C,2011ApJ...741...73V,2011Sci...333..203B,2011Sci...333..199L,2011Natur.476..425Z,2012MNRAS.420.2684C,2012ApJ...753...77C,2012Natur.485..217G}.
With more transient surveys~\cite{2009PASP..121.1334R,2004SPIE.5489...11K,2009arXiv0912.0201L}
beginning operation, the number of observed events should increase significantly,
making it important to understand the details of the events across a range of parameters.
For BHs with masses around $10^7$ to $10^8$ $M_{\odot}$, solar-type stars will
be tidally disrupted near the innermost stable circular orbit of the BH.
They will therefore be sensitive to strong-field effects including zoom-whirl type behavior
and the spin of the BH~\cite{Stone:2011mz,Kesden:2012qb}, which may be reflected in observations.

Numerous approaches have been applied to studying tidal disruptions. 
Analytical approximations include those based on Newtonian dynamics~\cite{1983A&A...121...97C,1988Natur.333..523R,1989IAUS..136..543P,1995MNRAS.275..498D,1999ApJ...514..180U,2009MNRAS.400.2070S}, 
Newtonian dynamics with relativistic corrections~\cite{1985MNRAS.212...57L,2010A&A...511A..80B,Stone:2012uk},
and incorporating aspects of Kerr geodesic motion~\cite{Kesden:2012qb}. There have also
been particle and grid-based simulations of these events utilizing 
Newtonian gravity~\cite{1982ApJ...263..377N,2012ApJ...757..134M,2012arXiv1206.2350G};
pseudopotentials to incorporate features of general relativity~\cite{1989ApJ...346L..13E,1997ApJ...479..164D,Hayasaki:2012ia,Hayasaki:2012yy}; or 
hydrodynamics on a fixed BH spacetime,
thus ignoring the self-gravity of the star~\cite{1993ApJ...410L..83L,2004ApJ...610..707B}.
In certain regimes, each of these methods is expected to decently approximate aspects
of the desired physics. However, there has yet to be a fully self-consistent calculation
within general relativity to investigate this, in particular for the case where disruption occurs near the innermost
stable orbit of the BH.
The details of the disruption process will depend on the interplay of the strong-field gravity of the black hole, 
the star's pressure, and the star's self-gravity, which is essentially Newtonian since $M_{\odot}/R_{\odot}~\sim2\times10^{-6}$.
The methods presented here allow us to perform general-relativistic hydrodynamic simulations that self-consistently 
combine all these components, and hence investigate their importance.   
As a demonstration, we present results from simulations
of the radial infall of a solar-type star into a BH, which can be easily compared to perturbative calculations.
We leave the study of the more astrophysically relevant parabolic orbits to future work.

In what follows we explain our method for subtracting background-solution truncation error and its implementation
in a general-relativistic hydrodynamics code.
We apply this method to simulating
the radial infall of a solar-type star into a supermassive BH, illustrating its efficiency and 
commenting on the tidal effects and resulting gravitational radiation.

\section{Computational methodology}
\label{comp_methods}
\subsection{Background error subtraction technique}

In this section we outline our background error subtraction technique (BEST), a method for altering the truncation error in cases where the system
can be written in terms of a known background solution, which satisfies the evolution equations on its own,
and a small perturbation.  The basic idea is straightforward.  Say we want to numerically 
find the solution $y(x,t)$ to
some evolution equation $\partial y/\partial t = \mathcal{F}$, where $\mathcal{F}$ is a nonlinear
operator.  We discretize $t$ as $t_n=n \Delta t$ and let $\Delta$ be a discrete evolution operator (e.g., a Runge-Kutta
time stepper) so that we can approximate the evolution as $y_{n+1}=\Delta(y_{n})$.     
Now consider the case where we can write 
$y(x,t)=\bar{y}(x,t)+\delta(x,t)$, where $\bar{y}$ is itself a known solution to the evolution equation and $|\delta| \ll |\bar{y}|$ in at least part of the domain.
In general,  even if $\delta(x,t)=0$, there will be truncation error from evolving $\bar{y}$. In fact, 
this error can be calculated exactly as $E_n = \Delta(\bar{y}(t=t_n))-\bar{y}(t=t_{n+1})$.  When evolving $y$, we can therefore explicitly subtract out the truncation error
from evolving only $\bar{y}$ at every time step, 
\begin{equation}
y_{n+1} = \Delta(y_n) - E_n.  
\label{tre_subtract}
\end{equation}
Since $E_n$ is converging to zero as $\Delta t \rightarrow 0$
at whatever order the numerical scheme converges, including this term does not change the overall order of convergence, nor
the continuum solution. 
However, where the truncation error
from evolving the background part of the solution dominates, including this term can reduce the magnitude of the truncation error since
the remaining error just comes from $\delta$ and its nonlinear interaction with $\bar{y}$.
Indeed, in the limit of vanishing $\delta$, we merely recover the exact solution $\bar{y}$.  
In the other limit, supposing $|\bar{y}|\ll |y|$, hence $\delta\approx y$,
the contribution from the $E_n$ term in Eq.~(\ref{tre_subtract}) will be negligible, and the 
solution from the unmodified numerical evolution scheme will be recovered. Though if this
were true in the entire domain, there would be no advantage to using this 
algorithm.

\subsection{Numerical implementation}

We apply the above method to evolving the Einstein equations in the generalized harmonic 
formulation~\cite{Pretorius:2004jg} where the dynamical variables are the metric and its time derivatives, 
$g_{ab}$ and $\partial_t g_{ab}$.  In general, evolution
equations can also be specified for the source functions $H^a:=\Box x^a$, though for simplicity here
we restrict ourselves to gauge choices where the source functions are specified as some function of the coordinates and metric variables.  
We consider cases where the metric is close to a known background solution and hence can be written as
$g_{ab} = \tilde{g}_{ab}+h_{ab}$, where $\tilde{g}_{ab}$ is the known background solution and $|h_{ab}| \ll |\tilde{g}_{ab}|$
(in at least part of the domain) 
and similarly for $\partial_t g_{ab}$.  In the example below we take $\tilde{g}_{ab}$ to be the metric of an isolated 
black hole in a moving frame, though this method will work for an arbitrary metric.

We use a version of the code described in~\cite{code_paper} to numerically evolve the Einstein-hydrodynamics equations 
with adaptive mesh refinement, modified by BEST. We note that whenever we 
interpolate, extrapolate, or apply numerical dissipation to the evolution variables, we do so to the quantities $h_{ab}$ and $\partial_t h_{ab}$.
From the viewpoint of the adaptive mesh refinement driver, these are treated as the dynamical variables.  We evolve the metric in time using fourth-order Runge-Kutta
and evolve the fluid variables using second-order Runge Kutta.  The fluid variables are evolved using high resolution shock-capturing
techniques as described in~\cite{code_paper} with the following modifications.  For the conserved fluid quantities we evolve 
$\tau:=-S_t/\alpha-S_i\beta^i-D$ (where ${D,S_a}$ are the conserved fluid quantities defined in~\cite{code_paper} and $\alpha$ and $\beta^i$
are the lapse and shift respectively) instead of $S_t$. This 
gives better results when the internal energy is small compared to the rest mass.
Additionally, when calculating the source terms in the fluid evolution equations that involve $\partial_a g_{bc}$,
we numerically compute $\partial_a h_{bc}$ and then add $\partial_a \tilde{g}_{bc}$.

From a programming standpoint, modifying a standard general-relativistic hydrodynamics code to implement BEST is straightforward as it only entails calling the time stepping function 
twice for every physical time step: once with the background 
solution $\tilde{g}_{ab}$ and all matter sources set to zero, and again with the full solution $g_{ab}$ and matter sources.
These results are then combined following Eq.~(\ref{tre_subtract}).
This will essentially double the computational expense of evolving the metric;
however, as seen below, the savings from not having to resolve
the background metric at the same level 
can more than make up for this.  If $\tilde{g}_{ab}$ is static
then it is only necessary to compute $E_n$ once for a given numerical grid.
This algorithm does not depend on the details of the particular numerical time stepper 
used nor the particular form of the background solution.
We also note that with this algorithm the level at which numerical round-off errors come in is still set by the magnitude 
of $g_{ab}$ and not by the magnitude of $h_{ab}$.

For the application considered in this paper, we use the axisymmetry of the problem to restrict our computational domain
to two spatial dimensions using a modified Cartoon method~\cite{Alcubierre:1999ab} as described in~\cite{Pretorius:2004jg}.  
However, the methods described here work equally well in three dimensions.
 
\subsection{Comoving frame}

For the application considered here we use a background solution that is a Galilean transformation of a static BH solution.
Specifically, we take an isolated BH solution in coordinates $\{\bar{t},\bar{x}^i\}$ and transform to the new coordinates
$\{t,x^i\}$ where $t=\bar{t}$ and $x^i=\bar{x}^i-p^i(\bar{t})$
where $p^i(\bar{t})$ is some specified function.  
Below we take $p^i$ to be the geodesic on the isolated BH 
spacetime with the same initial conditions as the star's center-of-mass.  This ensures that in the new
coordinates the star's center-of-mass will essentially be at coordinate rest.  
This is beneficial since the fluid sound speed $c_s$ is much smaller than the speed of light, and 
letting the star advect across the grid at speeds much greater than $c_s$ 
can lead to a loss of numerical accuracy (see~\cite{2012arXiv1206.2350G} and references therein).
For cases where the geodesic used to compute $p^i$ falls into the BH (as considered below) we transition
to a constant $p^i$ after the geodesic crosses the BH's horizon.

\section{Application}
\label{application}
\subsection{Setup}
As an application of BEST we consider a setup with a star of solar-type compaction $m/R_*=2\times 10^{-6}$ (where $m$ and $R_*$
are the mass and radius of the star, respectively) that falls radially into a black hole
of mass $M$.  The star is modeled as a perfect fluid with a $\Gamma=5/3$ equation of state.
We begin the star at a distance of $50M$ from the BH with the velocity of a geodesic falling from rest at infinity.  
The initial data is constructed 
by solving the constraint equations as described in~\cite{idsolve_paper}.
For the BH we begin with a harmonic solution~\cite{Cook:1997qc} and then apply a Galilean transformation
as described above to keep the star at approximately coordinate rest.
We evolve with the gauge choice $H^a=\tilde{\Box}(\tilde{x}^a)$ where all the quantities on the right hand side are
from the isolated (and Galilean-transformed) BH solution and hence are not functions of the dynamical variables. 
This ensures that the background solution does not undergo nontrivial gauge
dynamics during evolution.\footnote{In principle, any gauge condition which preserves the desired background solution
is allowed.  E.g., for a BSSN-puncture evolution, one could use the isotropic Schwarzschild solution with some 
variation of the $1+\rm{log}$ slicing and gamma-driver condition~\cite{Alcubierre:2002kk}.}
We consider mass ratios of $m/M=10^{-6}$ and $1.25\times 10^{-7}$.

For the $m/M=10^{-6}$ case, we use a grid setup with eight levels of mesh refinement (with 2:1 refinement ratio) covering 
the star's radius with approximately 50, 75, and 100 points for what we will refer to as 
the low, medium, and high resolutions runs, respectively.  
Unless otherwise specified, results below are from the high resolution runs with the other two
resolutions used to establish convergence.
For the $m/M=1.25\times10^{-7}$ case, we add additional refinement levels to achieve the same resolution 
covering the radius of the star.  As described in~\cite{code_paper}, during evolution the mesh refinement 
hierarchy is dynamically adjusted based on truncation error estimates.

\subsection{Comparison to not using BEST}
\begin{figure}
\begin{center}
\includegraphics[height = 2.8 in]{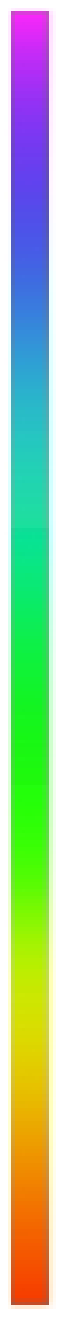}
\put(2,195){$10^{-7}$ }
\put(2,0){$10^{-12}$}
\hspace{0.32 in}
\includegraphics[width=2.8in,clip=true]{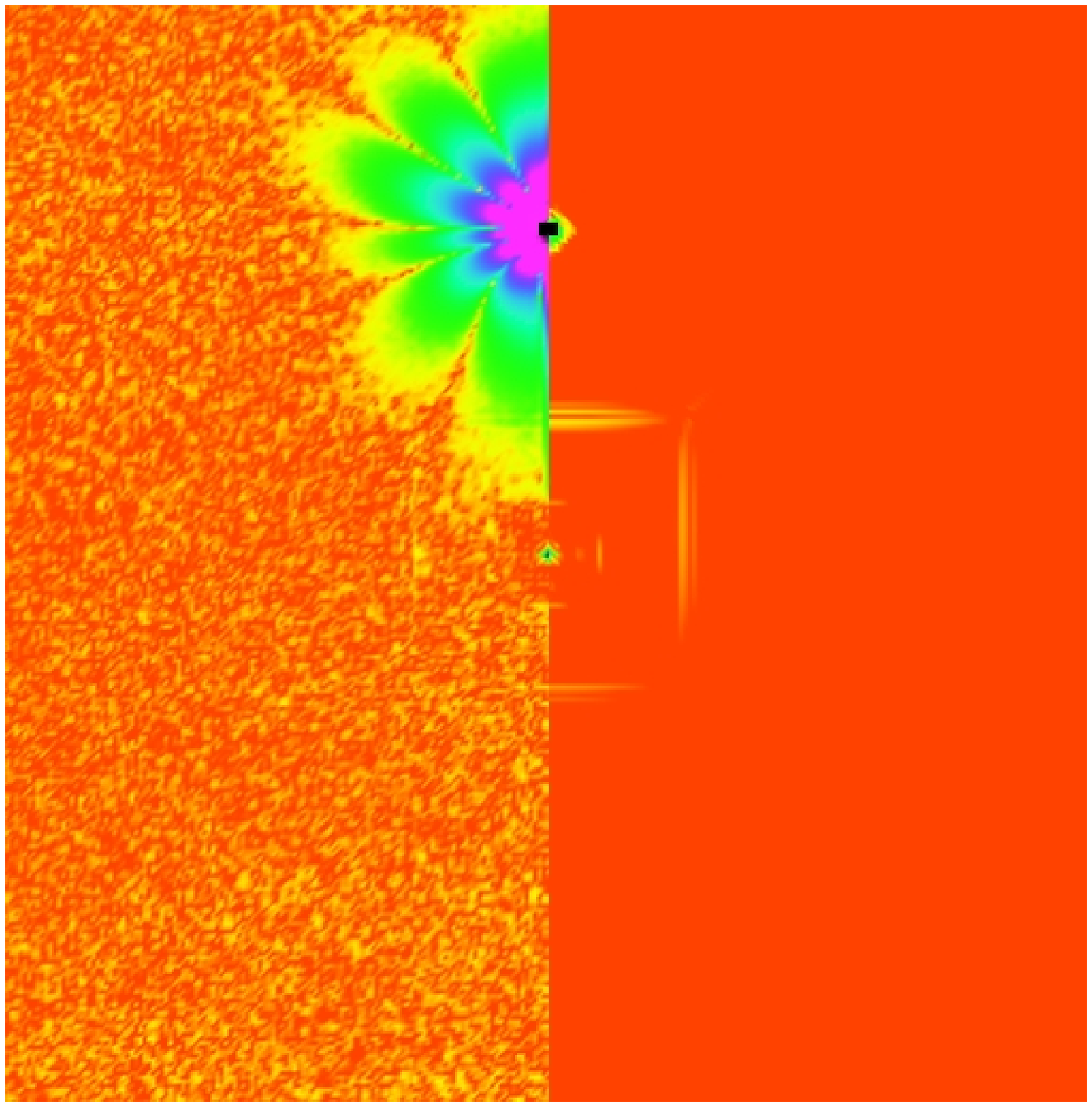}
\caption{
Comparison of the absolute magnitude of the truncation error in $g_{xx}$ without (left half) and with (right half)
the background subtraction algorithm.  Truncation error is calculated by comparing the quantity
after one coarse time step ($t\approx0.4M$) at lower resolution to the same quantity computed with four times the
resolution.  The inner $ [-100M,100M]\times[0, 100M]$ of the domain which is shown (with the $x$ axis in the vertical direction) 
is covered entirely by the second level of mesh refinement.  The star (center) is covered by 6 additional levels of refinement while
the BH (top) is not.
The color scale is logarithmic and is saturated in the left panel, which has a maximum of $\sim10^{-2}$.
\label{gb_xx_tre}
}
\end{center}
\end{figure}

In Fig.~\ref{gb_xx_tre} we illustrate the benefits of 
using the background error subtraction technique by plotting the truncation error in the
metric component $g_{xx}$ after one coarse time step with and without this technique
for the $m/M=10^{-6}$ case.
For this comparison, the same numerical grid at the low resolution is used and the 
evolution is carried out in exactly the same way except for the inclusion of the second
term in Eq.~(\ref{tre_subtract}) when taking a time step.
Since there is a lot of resolution concentrated on the star, 
at the initial separation the truncation error of the BH background solution is negligible
in the neighborhood of the star and the use of BEST does not make much difference.  
However, away from the star, and in particular near the 
BH, the truncation error from the background solution of an isolated BH moving across 
the grid is large.
The use of BEST makes a significant difference by obviating the need to use 
high resolution globally. 

Whereas in this example the BH is initially resolved at the same level as the wave zone
(six refinement levels fewer than the star),
in order to achieve the same level of truncation error near the BH after one coarse time step without BEST,
the BH must be covered with seven additional levels of refinement.  Using the total number of time steps that must be
taken at each point in the grid (where, since we use a refinement ratio of two, each successively finer refinement level takes twice as many steps to keep a fixed Courant factor) as an estimate of computational expense, 
the grid setup necessary without BEST 
is $\sim 40$ times more expensive (and would be $\sim 140$ times more expensive if our computational domain were three- instead of two-dimensional).  
This far outweighs the computational expense of computing the background error term when taking a 
time step, which will do no more than double the expense of taking a time step.   

\begin{figure}
\begin{center}
\includegraphics[width=3.4in,clip=true]{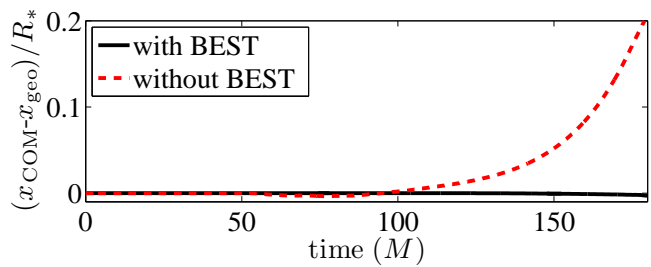}
\caption{
The distance of the star's center-of-mass from the equivalent geodesic for
$m/M=1.25\times 10^{-7}$ with and without the background subtraction algorithm
at low resolution. 
\label{star_com}
}
\end{center}
\end{figure}

We note that high accuracy is required to extract the gravitational wave signal
from this system (see Sec.~\ref{gwaves}) and when the evolution is performed without using BEST, 
even at the equivalent high resolution, truncation error completely dominates over the physical 
signal.  BEST makes little difference in modeling the star's self-gravity effects
noted in Sec.~\ref{selfgrav} (which is not surprising as the star is well resolved). 
However, the accumulation of truncation error from evolving without
BEST can cause the star's center-of-mass to drift from the geodesic path as shown in Fig.~\ref{star_com}. 

\subsection{Effects of self-gravity}
\label{selfgrav}
\begin{figure}
\begin{center}
\includegraphics[width=3.4in,clip=true]{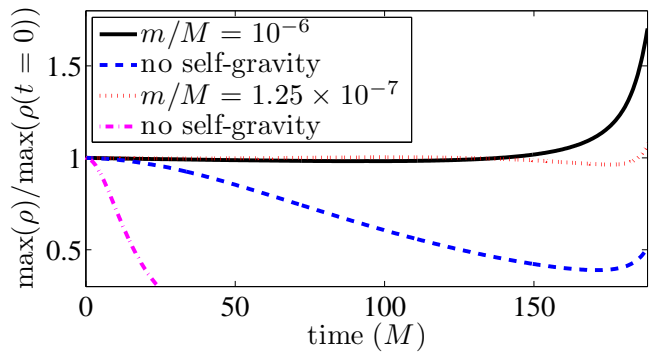}
\caption{
Normalized maximum rest density as a function of time with and without self-gravity for the
star for $m/M=10^{-6}$ and $m/M=1.25\times10^{-7}$. 
\label{max_rho}
}
\end{center}
\end{figure}

To demonstrate the importance of including the star's self-gravity in this calculation, 
we also consider simulations where we fix the metric to be that of the isolated BH.  
Without self-gravity to balance the star's pressure, it will expand outwards on 
timescales of $\sim R_*/c_s$.  In Fig.~\ref{max_rho} we show the maximum rest density
as a function of time with and without self-gravity.  For $m/M=10^{-6}$ the star's
central density drops by more than a factor of two before the star reaches the BH
(for this case $R_*/c_s\approx 370 M$ at the star's center).  
For $m/M=1.25\times10^{-7}$, as expected, this drop in density occurs approximately eight times 
faster in units scaled by the mass of the BH.  
With self-gravity, the star's central density remains essentially constant in both 
cases until the star gets close to the BH, at which point it increases.
Hence, simply calculating hydrodynamics on a fixed
spacetime background will not capture the correct physics.

\begin{figure}
\begin{center}
\includegraphics[width=3.4in,clip=true]{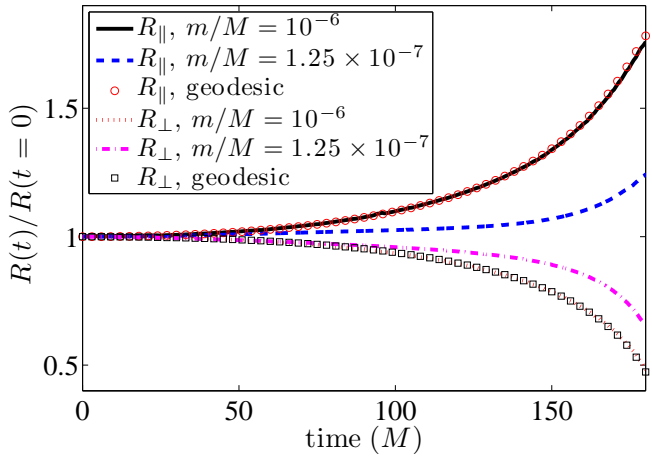}
\caption{
Normalized radius of the star perpendicular and parallel to the star's trajectory as a function of time
for $m/M=10^{-6}$ and $m/M=1.25\times10^{-7}$.  For comparison we also show the relative 
position of geodesics starting at corresponding points on the stellar surface
and with the same initial velocity as the star's center of mass. 
\label{star_radius}
}
\end{center}
\end{figure}

As the star falls into the BH, the star is stretched in the direction
parallel to its motion (i.e., the radial direction) and squeezed in the
perpendicular direction by the BH's tidal forces.  In Fig.~\ref{star_radius} 
we show the coordinate parallel and perpendicular radii of the $0.1\rho_c$ density contour
(where $\rho_c$ is the initial central density of the star) that initially contains $\approx90\%$
of the star's mass.
We compare this to the change in separation that two geodesics in the isolated BH 
spacetime would undergo if they had the same initial velocity and separation.
For $m/M=10^{-6}$, it seems that the combined effect of pressure and self-gravity
is small and the change in radii matches the geodesic calculation well. 
This is not surprising since the star begins at the nominal Newtonian tidal radius
of $r_{T} := R_* (M/m)^{1/3} = 50 M$.  For $m/M=1.25\times10^{-7}$ the tidal radius
is $r_T=12.5M$, and there is less of a change in the star's radii compared to freefall
at early times.  

\subsection{Gravitational waves}
\label{gwaves}
\begin{figure}
\begin{center}
\includegraphics[width=3.4in,clip=true]{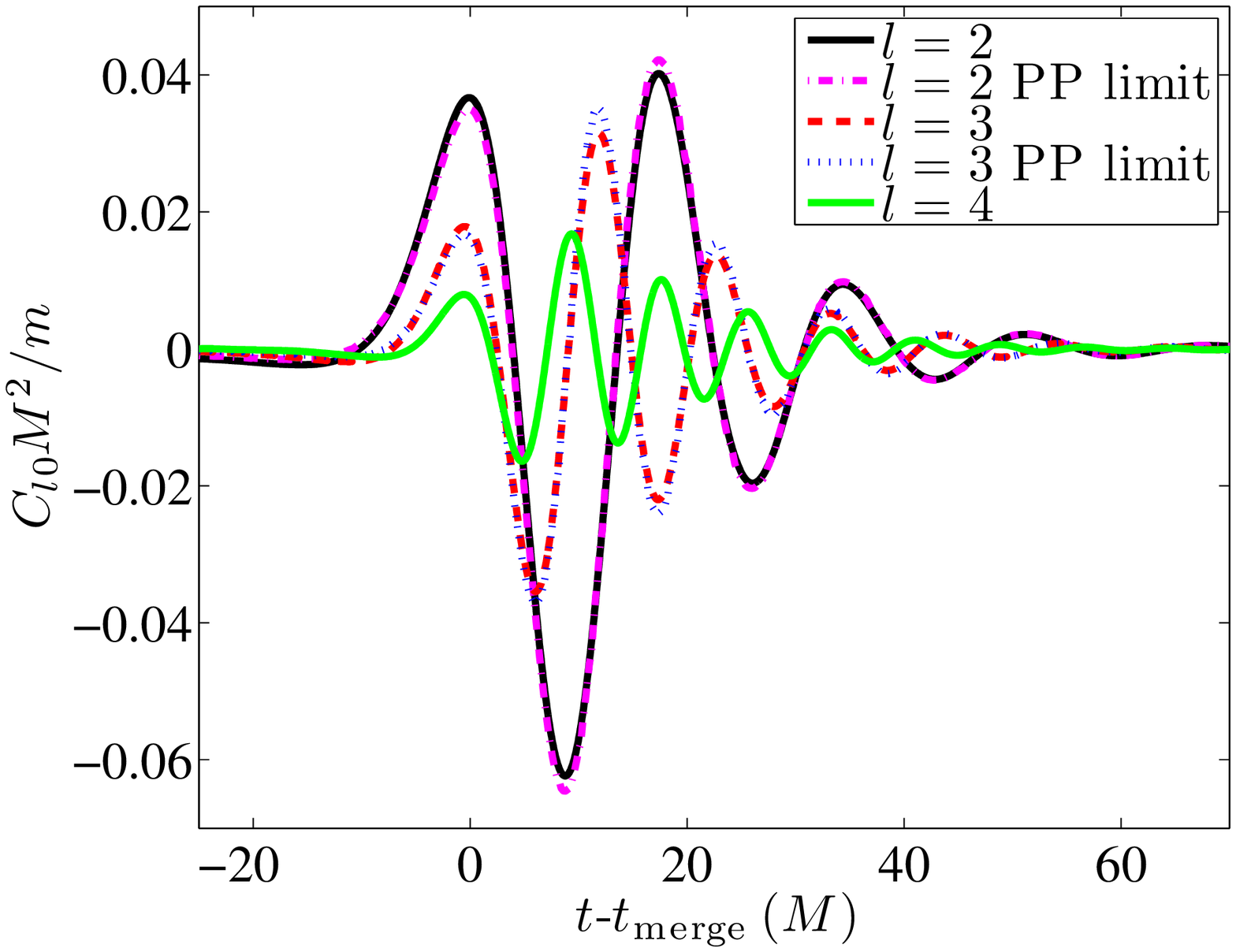}
\includegraphics[width=3.4in,clip=true]{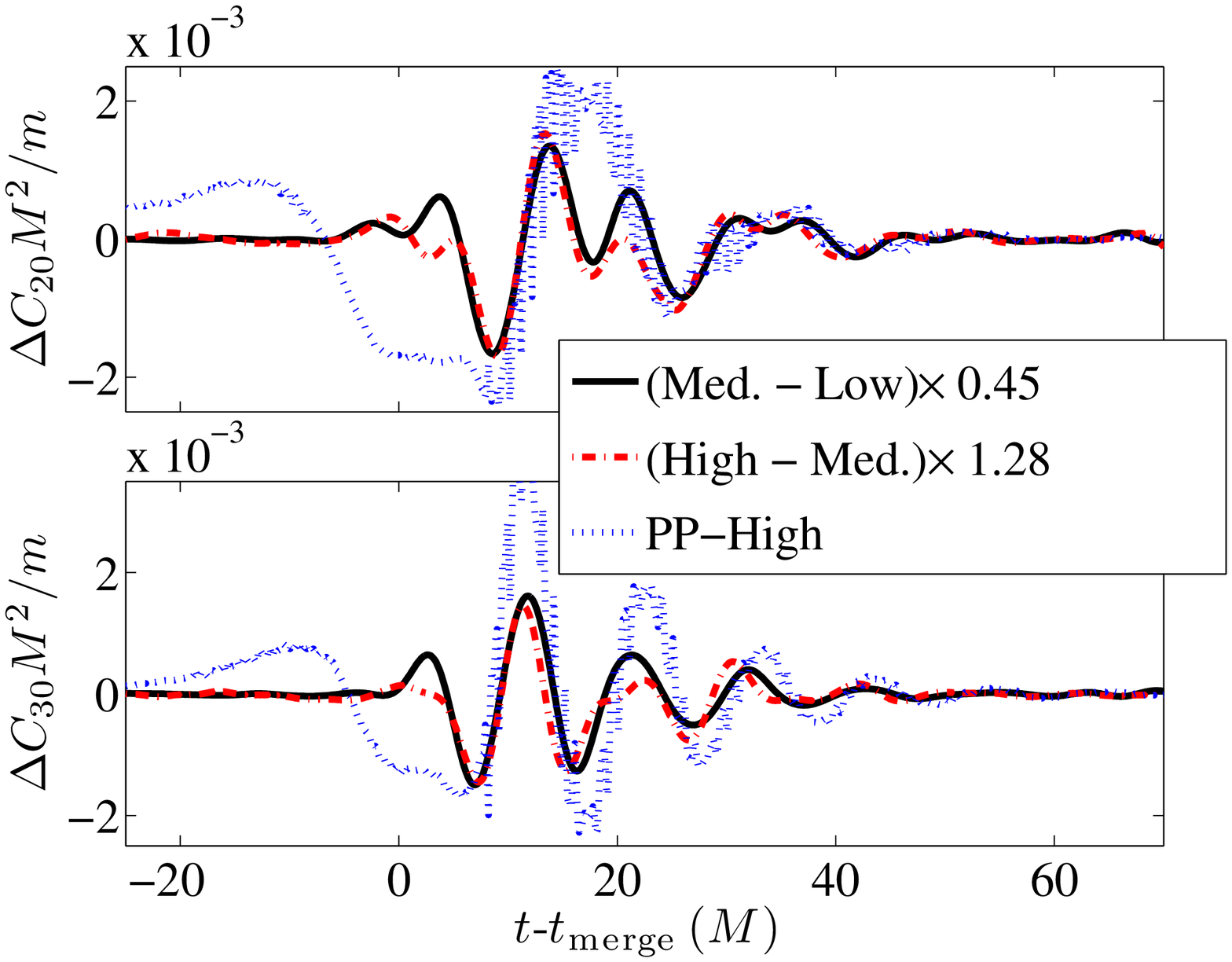}
\caption{
Gravitational wave signal from a star falling into a BH with $M=10^{^6}m$.
{\emph Top}: The first three spin-weight $-2$ spherical harmonics of $r\Psi_4$ 
as well as the first two harmonics as calculated using a point-particle approximation,
from~\cite{Sperhake:2011ik}, for comparison.
{\emph Bottom}: The difference between the $l=2$ and $l=3$ harmonics with 
resolution, scaled assuming second-order convergence, as well as the difference between
the highest resolution run and the point-particle calculation. 
\label{mbh1e6_gw}
}
\end{center}
\end{figure}

Since we are evolving the full spacetime metric, we can also self-consistently calculate the gravitational
wave signal.
In Fig.~\ref{mbh1e6_gw} we show the gravitational waves emitted from the star-BH interaction for the $m/M=10^{-6}$ case.
We plot spherical harmonics of the Newman-Penrose scalar multiplied by the extraction radius 
(because of the axisymmetry, only the $m=0$ components are nonzero).
The waveforms are shown multiplied by $M/m=10^6$, since in the point-particle limit this scaled quantity
is independent of the mass ratio. 
We also show the difference in the computed gravitational wave signal with resolution, which is consistent
with second-order convergence.

For comparison, we also show the gravitational wave signal of a point particle falling in a BH, which
was calculated in~\cite{Sperhake:2011ik} using BH perturbation theory~\cite{Davis:1971gg}.
Though at this mass ratio we are well within the perturbative regime, the star itself is not that close to a point mass since
$R_*=0.5M$.  Nevertheless, we find that our results are well matched by the point-particle results, and the difference
between the waveforms is comparable to the truncation error.
For the high resolution run, the total energy radiated is 0.0101 (0.0103) $m^2/M$, where the value in parentheses is 
the Richardson extrapolated value using all three resolutions and can be used to judge the error.
This is compared to 0.0104 $m^2/M$ for the point-particle result~\cite{Davis:1971gg}.

\section{Conclusion}
\label{conclusion}
We have presented a method, BEST, for more efficient solution of the Einstein 
equations in situations where the metric is dominated by a known background
solution.  We have demonstrated the utility of this method by applying 
it to the radial infall of a solar-type star into a supermassive black hole
and achieving $\sim 40$ decrease in the computational expense.
To our knowledge, this is the first computation within full general relativity
of the radial collision problem with such extreme mass ratios and relative
compaction between the two objects (upwards of $10^6$:1 and $10^5$:1, respectively).
We found that despite the comparable radius of the star and BH, and 
the importance of tidal forces in the star, the 
gravitational waveform from merger matches the point-particle
calculation to within the numerical error of a few percent.

The method outlined here is rather general and could be applied to many more problems.
An obvious extension, which we will address in future work, is to study 
tidal disruption of stars on parabolic orbits by supermassive BHs and
explore strong-field effects, including the spin of the BH.
This technique could also be used to more efficiently study other large-mass-ratio systems, 
such as binary BHs or a supermassive BH-neutron star merger, where both objects are strongly
self-gravitating, but the effect of the small object on the larger one is small.
Though the disparate length scales would still be computationally challenging,
there would be less need for high global resolution.
Other potential applications include simulating stellar-mass compact object binaries interacting 
in some strong-field background, such as near a supermassive BH, 
or possibly even studying cosmological systems
like nonlinear effects of fluctuations on a Friedmann-Robertson-Walker background. 

\acknowledgments
We thank the authors of~\cite{Sperhake:2011ik} for providing the point-particle waveforms shown
here. We thank Sean McWilliams and Branson Stephens for useful conversations. 
This research was supported by the NSF
Graduate Research Program under Grant No. DGE-0646086 (WE), NSF
Grant No. PHY-0745779, and the Simons Foundation (FP).
Simulations were run on the {\bf Orbital} cluster 
at Princeton University.

\bibliographystyle{h-physrev}
\bibliography{bhstar}

\end{document}